\documentclass[%
 reprint,
superscriptaddress,
 amsmath,amssymb,
 aps,
]{revtex4-1}
\usepackage{hyperref}
\usepackage{graphicx}
\usepackage{dcolumn}
\usepackage{bm}
\usepackage{textcomp}
\usepackage{subcaption}
\usepackage{xcolor}
\usepackage{hyperref}

\definecolor{c1}{HTML}{FF29B1}
\definecolor{c2}{HTML}{29FFC6}
\definecolor{c3}{HTML}{FFBB29}

\begin{document}

\preprint{APS/123-QED}

\title{Quantum Circuits for Toom-Cook Multiplication}

\author{Srijit Dutta}%
\affiliation{%
 Department of Computer Science and Engineering, IIT Bombay, India
}%
\author{Debjyoti Bhattacharjee}
\email{debjyoti001@e.ntu.edu.sg}
\affiliation{%
	School of Computer Science and Engineering, Nanyang Technological University, Singapore
}%
\author{Anupam Chattopadhyay}
\affiliation{%
 School of Computer Science and Engineering, Nanyang Technological University, Singapore
}%


\date{\today}

\begin{abstract}
\noindent In this paper, we report efficient quantum circuits for integer multiplication using Toom-Cook algorithm. By analyzing  the recursive tree structure of the algorithm, we obtained a bound on the count of Toffoli gates and qubits. These bounds are further improved by employing reversible pebble games through uncomputing the intermediate results. The asymptotic bounds for different performance metrics of the proposed quantum circuit are superior to the prior implementations of multiplier circuits using schoolbook and Karatsuba algorithms.
\end{abstract}

\maketitle


\section{\label{sec:intro}Introduction}
\noindent Quantum computing has gathered significant attention by solving certain problems much faster than any known classical algorithm. In contrast to Boolean logic, quantum bits~(qubits) not only represent the classical 0 and 1 states but also any complex combination or \textit{superposition} of both, leading to a significant speed-up in computing. The Deutsch-Jozsa algorithm~\cite{deutsch1992rapid} and Shor's factorization algorithm~\cite{shor1994algorithms} are well-known examples demonstrating the power of quantum computing. This capability gives rise to the bounded-error quantum polynomial time~(BQP) complexity class with an open quest among computer scientists and mathematicians to establish the exact relation between BQP and other complexity classes. In order to accelerate scientific computing using the capabilities of a quantum computer, efficient quantum circuits for basic mathematical functions are needed. {The efficiency of a quantum circuit is measured by lower computational space~(number of qubits) and lower computational time~(logical depth).  For fault-tolerant, error-protected quantum circuits to implement the quantum algorithms, it is projected that a large number of physical qubits are required for every logical qubit~\cite{campbell2017roads}. Naturally, potential solutions to reduce the number of logical qubits contribute to the overall efficiency of the quantum~circuit.}

Multiplication is one of the elementary mathematical operations of arithmetic. Fast long integer arithmetic is at the very core of many computer algebra systems. In quantum computing, apart from being used as a block in itself, integer multiplication is used as a sub-routine in many applications such as Shor's integer factorization algorithm and in Newton iterations for calculating many functions like the inverse~\cite{soeken2017design}.

In this paper, we present quantum implementation of the Toom-Cook multiplication algorithm~\cite{toom1963complexity,cook1969minimum}, which can attain better asymptotic complexity than simple schoolbook multiplication and the Karatsuba based integer multiplication~\cite{parent2017improved}. We further improve these bounds by analyzing pebble games on complete trees.

\section{\label{sec:prior}Prior Works}
\noindent The problem of multiplication in the quantum domain has been explored previously. For small numbers, the na\"ive schoolbook multiplication works best, with a runtime complexity $\mathcal{O}(n^2)$ that also translates  to the logical depth in a quantum circuit realization. Karatsuba multiplication, implemented in quantum circuits~\cite{parent2017improved}, is usually faster when the multiplicands are longer than $320-640$~bits, which also provides asymptotic improvement in terms of Toffoli cost and Toffoli depth over the schoolbook multiplication. However, the number of qubits required for Karatsuba-based quantum implementation is higher than the schoolbook multiplication. In the realm of quantum circuits, so far, the Sch\"onhage-Strassen method~(using Fast Fourier transform) and Toom-Cook multiplication algorithm are not reported, even though, it is known from the classical implementations that these algorithms result in better run time, when the operand size is much larger. We primarily focus on the Toom-Cook multiplication in this work. As reported in our results, this leads to significant savings of all the performance metrics for an efficient quantum circuit, clearly outperforming the prior works.

The family of Toom-Cook methods is an infinite set of polynomial algorithms~(Toom-$2.5$, Toom-3, Toom-4, \textit{etc})~\cite{knuth1997art}. Instead of using the more common Toom-3 implementation, we present the work with Toom-$2.5$ to avoid  a division by 3 required by the former. This leads to reduction in overall circuit costs, as quantum division is costlier in terms of Toffoli count and Toffoli depth than simple addition or shift operations. Most of the higher Toom implementations require a similar division by constants that are not multiples of 2. The implementation of such divisions incur higher quantum costs and therefore we avoid them.

When moving to quantum domains, gate sets need to go beyond classical to create the \emph{superposition} effect of the inputs. The standard universal quantum gate library that efficiently implements fault-tolerant quantum error correction codes is the Clifford+$T$ library~\cite{amy2014polynomial,fowler2009high}. In this library, the cost of implementing a $T$-gate is sufficiently high to customarily neglect the cost of other Clifford group gates, while determining the total cost of the quantum circuit. Therefore, the number of $T$-gates is a metric to judge the cost of a quantum circuit. Also, the number of qubits used in a quantum circuit is another important standard, since the current quantum technologies still struggle to achieve error free computation for large count of qubits.  A study of the space-time trade off can be performed~\cite{wille2014trading} using these two metrics. Another metric of importance is the \emph{T-depth}. {\emph T-depth} is defined as the number of \emph{T-stages} is a quantum circuit where each such stage consists of one or more \emph{T} or $\emph{T}^{\dagger}$ gates performed concurrently on separate qubits.  {It is important to note that an input circuit with continuous parameter gates (e.g. $z$-rotation gate $R_z(\theta)$ ) is decomposed using a set of discrete, basis gates, typically from the Clifford+T library. The exact number of Clifford+T gates needed for such a continuous parameter gate depends on the desired accuracy, and the discrete gate set provides only an approximation. In the context of the current work, we consider T-count and T-depth of the circuit to be proportional to the Toffoli-count and Toffoli-depth respectively, by following the Toffoli decomposition proposed in~\cite{abdessaied2016technology}}.
%
%
%

\section{\label{sec:method}Toom-Cook Multiplier}
\noindent Given two large integers $n_1$ and $n_2$, the Toom-Cook algorithm splits them into $k$ smaller parts  of length $l$. The multiplication sub-operations are then computed recursively using Toom-Cook multiplication again, till we are able to apply another algorithm on it for the last stage of recursion, or until the desired multiplier is reached. The input numbers are divided into limbs of a given size, each in the form of polynomial, and the limb size is used as radix. Instead of multiplying the obtained polynomials directly, they are evaluated at a set of points and the values multiplied together at those points. Based on the products obtained at those points, the product polynomial is computed by interpolation. The final result is then obtained by substituting the radix.

In general, Toom-$k$ runs in $\Theta(c(k)n^e)$, where $n$ denotes input size,  $k$ is the number of parts that the input operand is decomposed into and \mbox{$e = \log_{k}{(2k-1)}$}. $c(k)$ is the  time spent on auxiliary additions and multiplications by small constants. The Karatsuba algorithm~\cite{karatsuba1963multiplication} is a special case of Toom-Cook multiplication~(Toom-2), where the input operand is split into two smaller ones. It reduces 4 multiplications to 3 and so operates at $\Theta(n^{\log_{2}{3}})$. In general, Toom-$k$ reduces $k^2$~multiplications to $2k-1$~ordinary long multiplication~(equivalent to Toom-$1$) with complexity~$\Theta(n^2)$.

\subsection{Implementation Details}
\noindent Let $x$ and $y$ be two $n$~bit numbers.  To proceed with Toom-$2.5$ algorithm, we first decompose $x$ and $y$ into two and  three parts respectively. Express $x = x_12^i + x_0$ and $y = y_22^{2i}  +  y_12^i + y_0$ with $i \geq 1$. Typically $i$ is chosen as $max\Big\{ \left\lfloor \frac{\lceil\log_{2}{x}\rceil}{k}\right\rfloor,\left\lfloor \frac{\lceil\log_{2}{y}\rceil}{k}\right\rfloor\Big\}$, where $k=2.5$ in our case. We define the following four product terms~:
\begin{align}
P &= x_0y_0,\\
Q &=  (x_0+x_1)(y_0+y_1+y_2),\\
R &= (x_0-x_1)(y_0-y_1+y_2),\\
S &=  x_1y_2
\end{align}
Then, the product~$xy$  is evaluated as :-
\begin{align}
xy &= A2^{3i} + B2^{2i} + C2^i + D\\
A &= S \\
B &= -P  +  \frac{1}{2}Q + \frac{1}{2}R\\
C &= \frac{1}{2}Q - \frac{1}{2}R -  S\\
D &= P
\end{align}
Note that only 4 multiplications are required for  computation of the product. Also, each of these multiplications consists of numbers of size smaller than the original problem size~(bit-width). Each smaller multiplication is between one number of bit-width $n/2$ and another of bit-width $n/3$. Since this method is  applied in recurrence the second time for our analysis, we consider that the smaller limbs formed from the number which was split into 3 parts originally is now split into 2 parts and vice-versa. So after two steps, we get 16~smaller problems of size~$n/6$ each. Thus, we obtain the basic recurrence for the number of steps~$T(n)$.
\begin{equation}\label{eq:recursion}
T(n) = 16 T(\frac{n}{6})
\end{equation}  
All additions~(the intermediate ones as well as the final ones) are performed by separate adders which have bounded cost.

\subsection{Gate count Analysis}
\noindent For gate count analysis, we consider only the Toffoli count required by the quantum circuit or sub-circuit. This is because the other used gates~(NOTs, CNOTs) do not contribute to the $T$-count of the circuit, considering the Clifford+$T$ library. The designed circuit maps \mbox{$(x,y,0,0) \mapsto (x,y,g,xy)$}, where $g$ denotes some garbage output resulting as computation of $A, B, C$ and $D$. The product is copied after the computation and the circuit is then run backwards~({\em uncomputed}) to set the garbage outputs back to~$0$. 

 In our circuit implementation, the Cuccaro adder is used~\cite{cuccaro2004new}. For addition of two $n$~bit numbers, Cuccaro adder requires $2n-1$~Toffoli gates. It is also established that the cost $A_n$, for an in-place adder adding two $n$~bit numbers, is bounded by $2n$~Toffoli gates. 
 
 Let $T_{n,n}$ denote the multiplication call to Toom-$2.5$ circuit for calculating product to two $n$~bit numbers and $TC_{n}$ denote the number of Toffoli gates required for implementing $T_{n,n}$. First, we need to calculate $P, Q, R$ and $S$. This requires 4 recursive calls to $T_{\frac{n}{2},\frac{n}{3}}$. 
For calculating  the intermediate sums required as input for $T_{\frac{n}{2},\frac{n}{3}}$, we need four $n/2$~bit adders and six $n/3$~bit adders. This also includes uncomputation of the intermediate garbage  results, i.e., the qubits used for storage of intermediate results are returned to their initial states. 
The output of each $T_{\frac{n}{2},\frac{n}{3}}$ is a $5n/6$~bit number. Finally, for computing $A, B, C$ and $D$, four $5n/6$~adders are required. As already stated earlier in evaluation of $T_{\frac{n}{2},\frac{n}{3}}$ we assume that the $n/2$~bit number is split into 3 parts and vice versa. By performing similar analysis, we get evaluation of $T_{\frac{n}{6},\frac{n}{6}}$, in terms of which the recursive relation is provided.  
\begin{align}
 TC_n &= 16TC_{n/6} + 40A_{n/6} + 22A_{n/3} + 4A_{n/2} + 4A_{5n/6} \\
  &= 16^{log_6n}TC_1 + 40(A_{\frac{n}{6}} + 16A_{\frac{n}{36}}+\dots)  \nonumber\\  &+ 22(A_{\frac{n}{3}} +  16A_{\frac{n}{18}}+\dots) +  4(A_{\frac{n}{2}} + 16A_{\frac{n}{12}}+\dots) \nonumber\\ &+ 4(A_{\frac{5n}{6}} + 16A_{\frac{5n}{36}}+\dots)
\end{align}
The base case is the multiplication of two $1$~bit numbers which can be done by a Toffoli gate. Therefore, \mbox{$TC_{1}$ = 1}. Each of the summation of the adder gate counts have log$_6n$ terms. On evaluating the summations using geometric progression and doubling the cost to account for the aforementioned uncomputation,  we get :-
\begin{align}
TC_n &=  2\big( 16^{log_{6}n} + 23.2n\big[\big( \frac{16}{6}\big)^{\log_6n}-1\big]\big)\\
&= 2n^{\log_{6}16} + 46.4n( n^{\log_{6}(16/6)}-1)
\leq 49n^{\log_{6}16}
\end{align}
 Note that all operations used in the circuit design are implemented using only adders and shifts, without any separate multiplication/division blocks.
\begin{center}
	\begin{figure}[t]
		\includegraphics[height = 5cm]{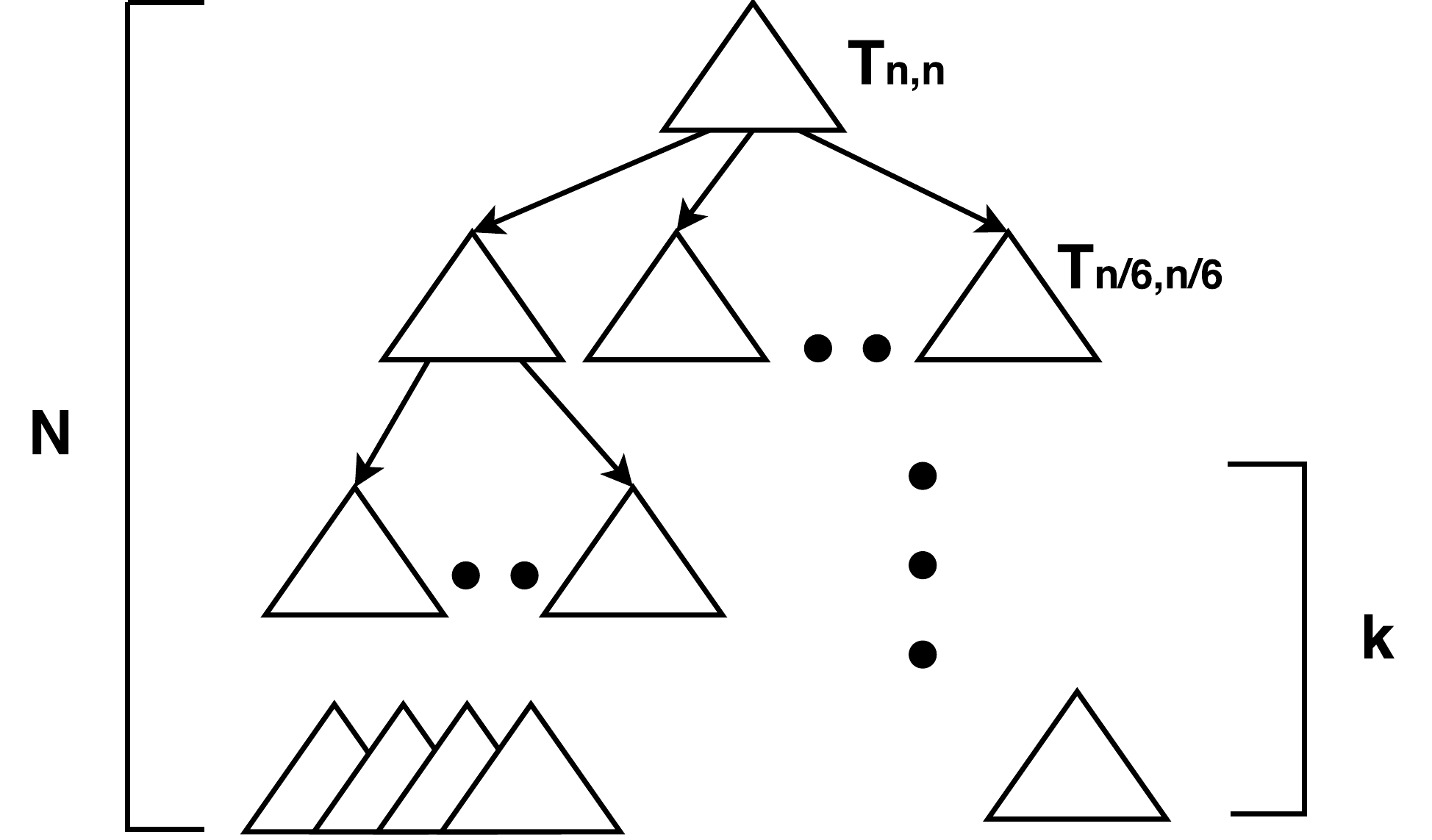}
		\caption{\em Recursion tree structure of the Toom-$2.5$~implementation.}
		\label{fig:tree}
	\end{figure}
\end{center}
\subsection{Space-Time Trade-offs}
\noindent The recursive nature of the problem gives rise to an inherent tree structure as shown in Fig.~\ref{fig:tree}. The size of a node is representative of the problem size at that level. For example, the root level denotes the complete problem~($n$~bit multiplication). According to the recursion presented in Equation~(\ref{eq:recursion}), each node will have $16$~children nodes denoting a smaller problem~($n/6$~bit multiplication).
For the Toom-$2.5$ circuit with an input of size $n$ at any level~$x$ in the tree, there are $16^{x}$ nodes of size $n6^{-x}$ each for a total cost of $n\big(\frac{16}{6}\big)^{x}$ at level $x$~(level numbering starting at $0$ from root). So, the space~cost~$Q_{orig}$ of the complete tree is 
\begin{align}
Q_{orig} &= n\sum_{0}^{N-1} \Big(\frac{16}{6}\Big)^{x} \\
&= n\frac{(16/6)^{\log_{6}{n}}-1}{(16/6)-1} \\
&= \mathcal{O}(n(8/3)^{\log_{6}{n}}) = \mathcal{O}(n^{1+\log_{6}{(8/3)}}) \\
&\approx \mathcal{O}(n^{1.547})
\end{align}
 where tree height, $N=\log_{6}{n}$. 

	\begin{figure*}[ht]
		\centering
		\includegraphics[height = 8cm]{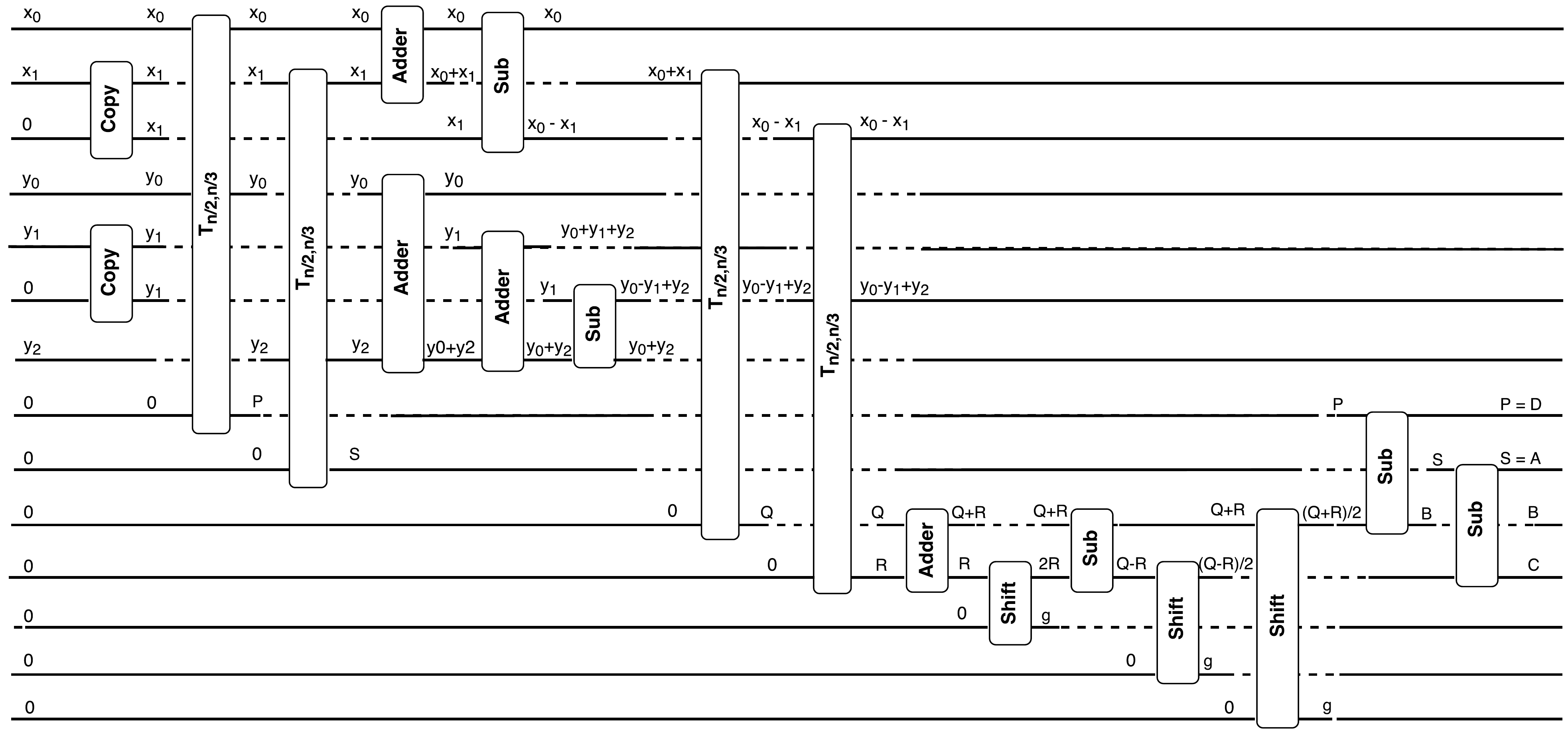}
		\caption{\label{fig:circuit} \em{The quantum circuit for computing integer multiplication result using Toom-$2.5$ algorithm. The compute blocks are then run backwards~(uncomputed) to set the garbage outputs~($g$) back to 0~(not shown in the figure).}}
	\end{figure*}

The reversible pebble game~\cite{bennett1989time} is a combinatorial game played on rooted Directed Acyclic Graphs~(DAGs). Each pebble represents some amount of space. The rules are similar to those used in the pebble game to model irreversible computation except that we simply cannot remove pebbles by reversibility constraint. There is a reverse computation for each corresponding computation performed, implying that during the game, the pebbles may still be removed but it is subject to the same conditions as applied during placing the pebbles. We use this reversible game to obtain better asymptotic bounds in the number of qubits~(space) to implement the Toom~$2.5$~algorithm.

We want to find a level in the recursion tree such that the size of each node's sub-tree is approximately equal to the sum of the size of all nodes at that level chosen and above. Once all the nodes in the chosen level have been computed, we uncompute all the sub-trees below it. This is performed to minimize space --- the size of these sub-trees is chosen to be approximately equal to the remaining size of the tree above them. Let the required height be $k$ from the leaves of the tree. The cost of all height $k$ sub-trees is $$ n\sum_{N-k}^{N-1} \Big(\frac{16}{6}\Big)^{x}$$
Therefore, cost of a single height $k$ sub-tree is  $$ \frac{n}{16^{N-k}}\sum_{N-k}^{N-1} \Big(\frac{16}{6}\Big)^{x}  =\frac{n}{6^{N-k}}\sum_{0}^{k-1} \Big(\frac{16}{6}\Big)^{x}  $$
We want this to equal the cost of all nodes above the $k^{th}$~level. 
\begin{equation}
n \sum_{0}^{N-k-1} \Big(\frac{16}{6}\Big)^{x} = \frac{n}{6^{N-k}}\sum_{0}^{k-1} \Big(\frac{16}{6}\Big)^{x}
\end{equation}
Simplifying we obtain a bound that $k \leq \frac{N}{2-\log_{16}{6}}$. This is since $k \leq N$ and $ \Big(\frac{16}{6}\Big)^{N-k} \geq \frac{16^k}{6^N}$.
Using the above technique, the qubit count is now optimized and bounded by $Q_{opti}$.
\begin{equation}
Q_{opti} = \mathcal{O}\Big( n\Big(\frac{8}{3}\Big)^{ \frac{1}{2-\log_{16}{6}} \log_{6}{n}}\Big) \approx \mathcal{O}(n^{1.404})
\end{equation}
The time complexity of a quantum circuit is effectively equal to the depth of the circuit in terms of Toffoli gates. Each node in the computation tree shown in Fig.~\ref{fig:tree} at level $k$, must be computed sequentially. At the $k^{th}$ level, the number of sub-trees $ST_k$ and corresponding depth $D_k$ is defined as follows.
\begin{align}
ST_k &= 16^{\big(1-{\frac{\log{16}}{2\log{16}-\log{6}}\big)\log_{6}{n}}} \\
D_k &= \frac{n}{6^{\big(1-{\frac{\log{16}}{2\log{16}-\log{6}}\big)\log_{6}{n}}}}\\
ST_k*D_k &= n\Big(\frac{8}{3}\Big)^{ \big(1- \frac{\log{16}}{2\log{16}-\log{6}}\big) \log_{6}{n}}\approx n^{1.143}
\end{align}
The product $ST_k*D_k$ gives an overall depth for computing the entire $k^{th}$ level of the recursion tree.

The method proposed above is most efficient if both the numbers to be multiplied are approximately of the same bit-width. In case one of them is much bigger than the other, it is better if the bigger number is repeatedly divided into 3 parts in each turn, until the smaller parts of both the numbers are roughly the same size. Following this method, the asymptotic computational complexity can be shown to be more efficient than that of the \textit{alternating} Toom-$2.5$ method adopted.

 The circuit of the described implementation is shown in Fig.~\ref{fig:circuit}. It describes the circuit for $T_{n,n}$ that multiples $x$~(decomposed into $x_0,x_1$) and $y$~(decomposed into $y_0,y_1,y_2$). All symbols and variables mentioned hold the same meanings as described in the analysis above. The adder, subtractor and shifting blocks are represented as `Adder', `Sub' and `Shift' respectively. The $T_{\frac{n}{2},\frac{n}{3}}$~blocks denote Toom-$2.5$ sub-circuits of smaller bit-width. 
\begin{table}[ht]
	\centering
	\caption{\em{Asymptotic performance analysis of the quantum implementation of various multiplication methods.}}
	{ 
	\begin{tabular}{ l l l l  }
		\hline\hline
		\textbf{Method} & \multicolumn{1}{l}{\textbf{QC}} & \multicolumn{1}{l}{\textbf{TC}} & \multicolumn{1}{l}{\textbf{TD}} \\
		\hline
		Na\"ive~\cite{parent2017improved}  &  ${\mathcal{O}(n)}$ &$\mathcal{O}(n^2)$ & $\mathcal{O}(n^2)$ \\
		Na\"ive Improved.~\cite{draper2004logarithmic} & $\mathcal{O}(n)$ & ${\mathcal{O}(n^2)}$ & ${\mathcal{O}(n\log n)}$ \\
		Karatsuba~\cite{parent2017improved} & $\mathcal{O}(n^{1.427})$ &$\mathcal{O}(n^{\log_23})$ & $\mathcal{O}(n^{1.158})$ \\
		\textit{Toom-$2.5$} & $\mathbf{\mathcal{O}(n^{1.404})}$ &$\mathbf{\mathcal{O}(n^{log_616})}$ & $\mathbf{\mathcal{O}(n^{1.143})}$ \\
		Const. Mult.~\cite{pavlidis2012fast} & $\mathcal{O}(n)$ & $\mathcal{O}(n^{2})$ & $\mathcal{O}(n)$ \\
		\hline\hline
		\multicolumn{4}{l}{QC: Qubit count, TC: Toffoli count, TD: Toffoli depth} \\
	\end{tabular}
}
	\label{table:1}
	%
	\caption{\em{Cost of quantum implementation of multiplication.}}
	\centering
	{\scriptsize
		\begin{tabular}{ l l l l  }
			\hline\hline
			\textbf{Method} & \textbf{QC} & \textbf{TC} & \textbf{TD} \\
			\hline
			Na\"ive~\cite{parent2017improved}  &  $4n+1$ &$4n^2-3n$ & $4n^2-4n+1$ \\
			Karatsuba~\cite{parent2017improved} & $ n\Big(\frac{3}{2}\Big)^{ \frac{ \log_{2}{n}}{2-\log_{3}{2}}}$ &$ 42n^{\log_{2}3}$ & $n\Big(\frac{3}{2}\Big)^{ \big(1- \frac{1}{2-\log_{3}2}\big) \log_{2}{n}}$ \\
			\textit{Toom-$2.5$} & $ n\Big(\frac{8}{3}\Big)^{ \frac{ \log_{6}{n}}{2-\log_{16}{6}}}$ &$ 49n^{\log_{6}16}$ & $n\Big(\frac{8}{3}\Big)^{ \big(1- \frac{1}{2-\log_{16}6}\big) \log_{6}{n}}$ \\
			Const. Mult~\cite{pavlidis2012fast} & $3n+1$ & $ 4n(n+1)$ & $8n$ \\
			\hline\hline 
			
		\end{tabular}
	}
	\label{table:impl}
\end{table}

\begin{figure*}[ht]
	\centering
	\begin{subfigure}[t]{5.5cm}
		\centering
		\includegraphics[width=\textwidth]{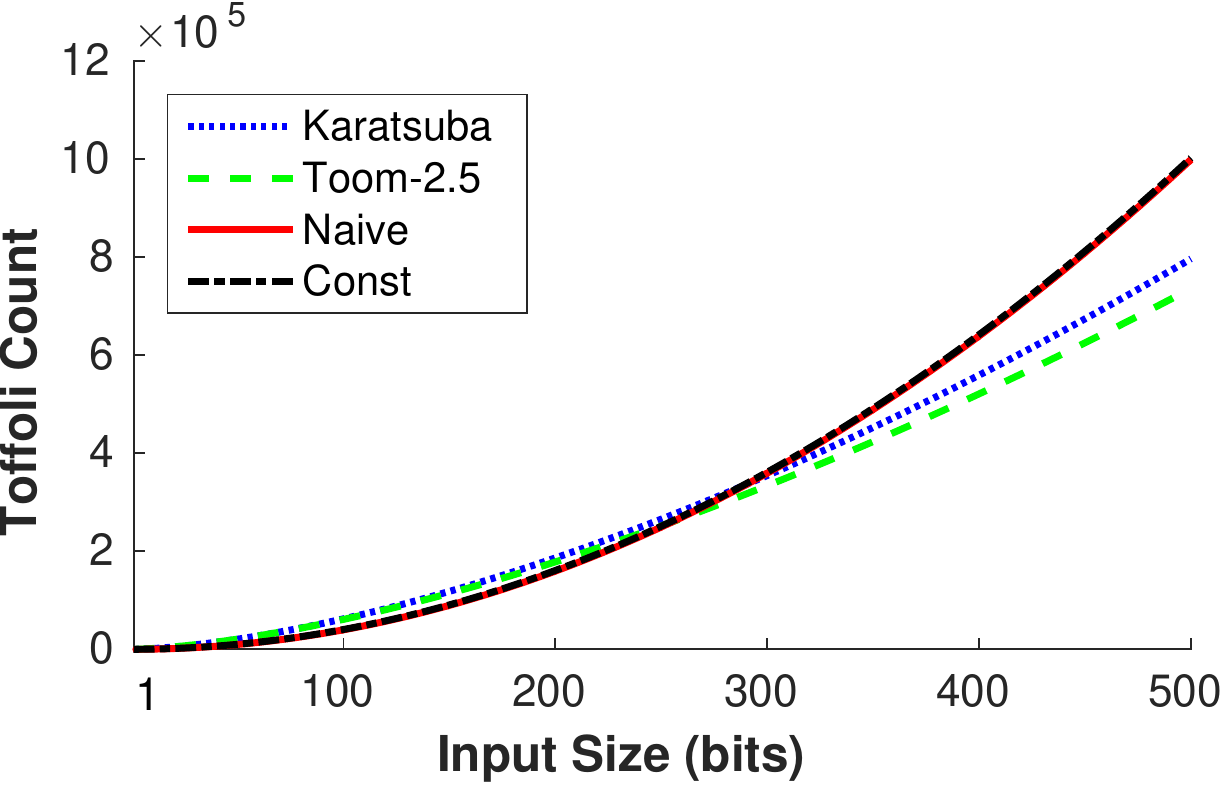}
		\caption{}
		\label{fig:tofgraph}
	\end{subfigure}
	\begin{subfigure}[t]{5.5cm}
	\centering
	\includegraphics[width=\textwidth]{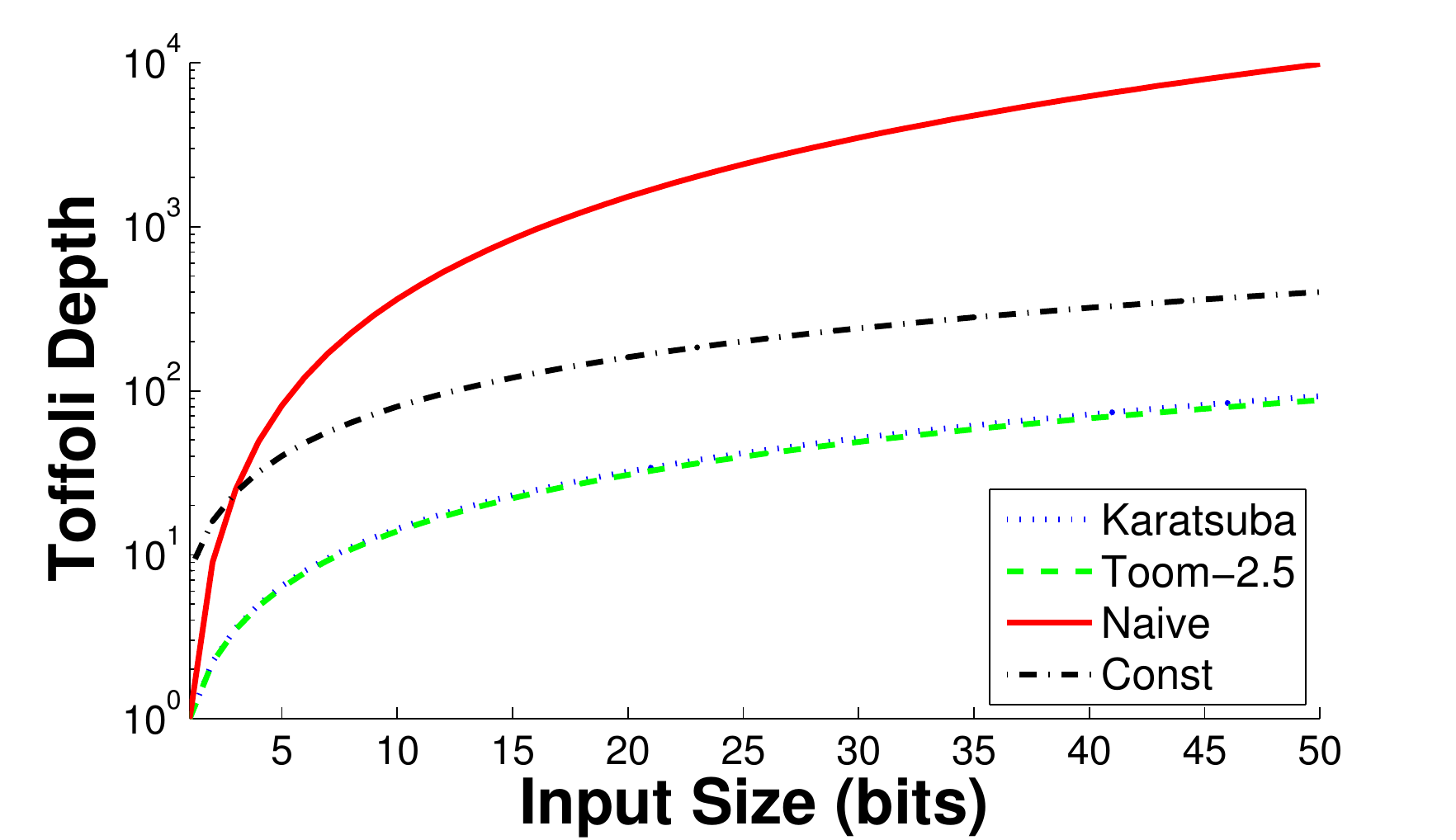}
	\caption{}
	\label{fig:depgraph}
	
\end{subfigure}
	\begin{subfigure}[t]{5.5cm}
		\centering
		\includegraphics[width=\textwidth]{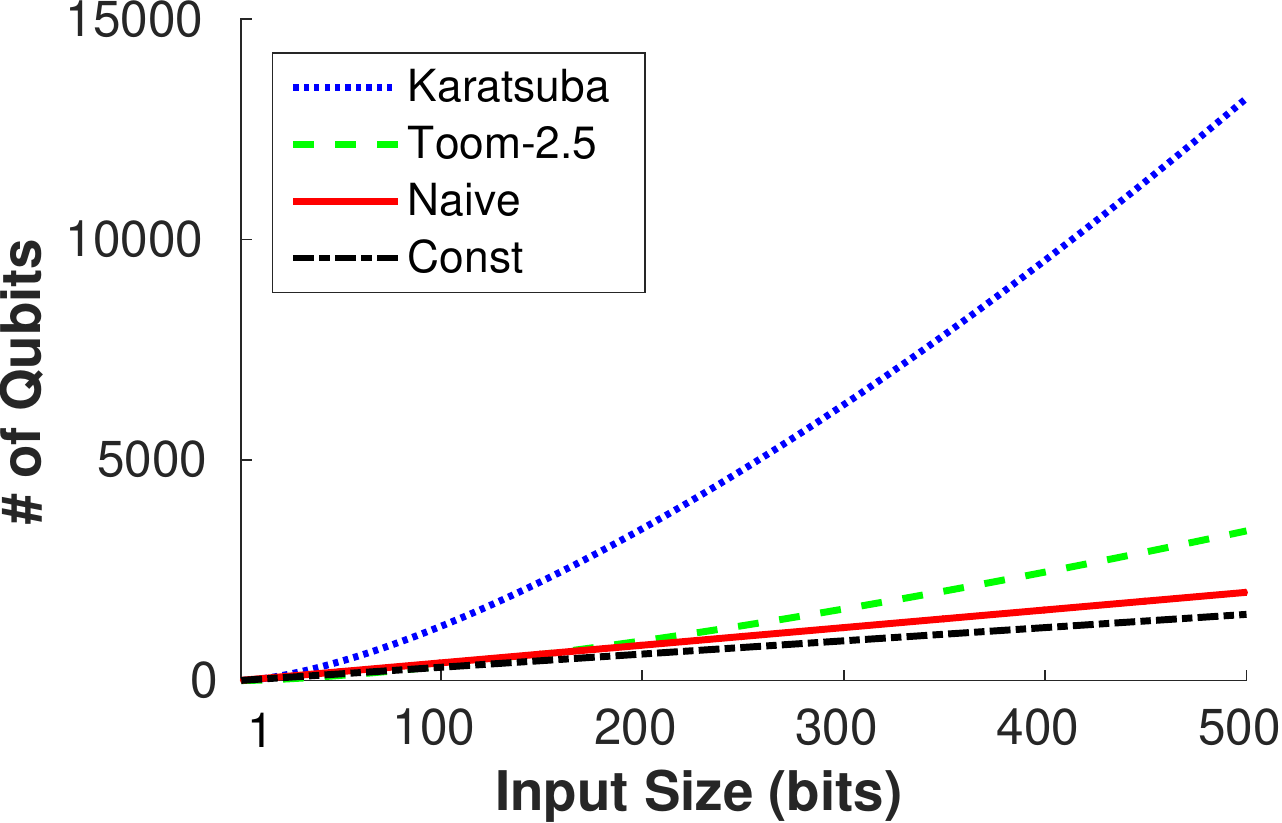}
		\caption{}
		\label{fig:qubitgraph}
	\end{subfigure}

	\caption{\em{Comparison of the quantum multiplier implementations based on : \subref{fig:tofgraph} Toffoli Count, \subref{fig:depgraph}~Toffoli~Depth} and \subref{fig:qubitgraph}~\#Qubits.}
	\label{fig}
\end{figure*}
\section{Results and discussions}\label{sec:exp}
\noindent Table~I presents the asymptotic results of implementation of various multiplication methods while Table~II provides the exact constants involved.
	The na\"ive multiplication method suggested in~\cite{parent2017improved} allows implementation with the lowest number of qubits asymptotically but fares badly in terms of Toffoli count and depth.
 
	In~\cite{draper2004logarithmic}, the implementation of  logarithmic depth adders have been provided.
The na\"ive~(shift-add) multiplier can be improved in depth by using the logarithmic depth adder as a submodule. The $n$-bit adder has a depth of order $\mathcal{O}(\log{n})$, thus the multiplier shall have a depth of $\mathcal{O}(n\log{n})$. However, for both `in place' and `out of place' adders described in~\cite{draper2004logarithmic},  extra ancilla are required for intermediate computation. Also, the Toffoli count is greater compared to the Cuccaro Adder. Thus, the multiplier developed by extension though optimized in depth, will have greater asymptotic Toffoli and qubit count, equal to $\mathcal{O}(n^2)$. In Table~I, we provide the asymptotic complexity of such a multiplier. However, in the absence of an explicit design, we are unable to provide the exact constants involved in the cost metrics and hence Na\"ive Improved method mentioned in Table~I is excluded from Table~II.
Toom-$2.5$ requires less number of qubits than Karatsuba~\cite{parent2017improved}. Toom-$2.5$ outperforms both the na\"ive and Karatsuba methods in terms of Toffoli count as well as Toffoli depth, highlighting the efficiency of the proposed method.

%
 Pavlidis et al.~\cite{pavlidis2012fast} presented a depth optimized multiplier, for multiplication by a constant only. Therefore, it is unfair to be directly compared with our implementation and the Karatsuba multiplication implementations presented in~\cite{parent2017improved}. It has a Toffoli depth of $8n$, a cost of $4n(n+1)$ and qubit count of $3n+1$.

{The Clifford+T quantum gate library has garnered 
much  interest  in  the  implementation  of  fault-tolerant  quantum  circuits~\cite{weinstein2013non}. As mentioned in~\cite{maslov2016optimal,shende2008cnot}, the cost of Toffoli gate is higher compared to the NOT and CNOT gates. The Toffoli gate may be decomposed using Clifford+$T$-gates, which makes cost metrics associated with Toffoli gates important. Therefore, Toffoli count and Toffoli depth are used as the performance metrics to begin with. The cost of mapping a Toffoli gate to the Clifford+T fault tolerant library is upper bounded by { $7\times$ Toffoli count} and {$3\times$ Toffoli depth} ~\cite{abdessaied2016technology}. Therefore, fault tolerant implementation of the proposed multiplication method would have at most $7\times$ Toffoli count and $3\times$ Toffoli depth of the values mentioned in Table~I. It is further possible to improve these values by optimization techniques proposed in~\cite{abdessaied2016technology,amy2014polynomial}.}
%
%
%
%
%

Fig.~\ref{fig:tofgraph} presents a comparison of the Toffoli count required by the various methods for variation in the bit-width of the inputs. The na\"ive multiplication method performs better in terms of total Toffoli cost at smaller input sizes~($<300$ bits), but is outperformed by the Karatsuba and Toom algorithms at higher bit-widths.
 
 Fig.~\ref{fig:qubitgraph} shows the variation in the qubit requirements by the different implementations across a range of input sizes. In this case the shift and add method~(na\"ive) outperforms both the recursive algorithms as it increases linearly. However, this low space requirement leads to a higher depth as demonstrated, in Fig.~\ref{fig:depgraph} in a logarithmic scale. Both Toom-$2.5$ and the Karatsuba implementations perform much better in this respect.

\begin{center}
	\begin{figure}[t]
		\includegraphics[height = 4cm]{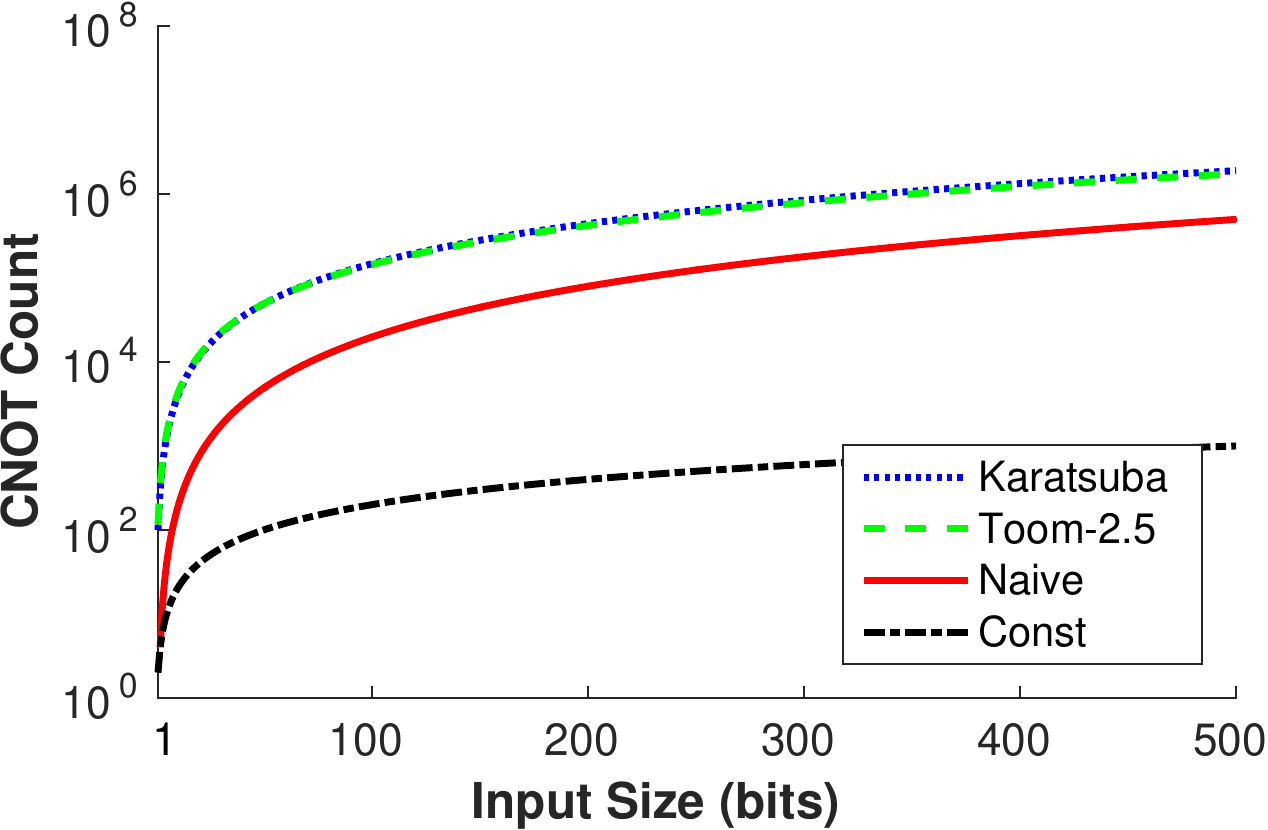}
		\caption{\label{fig:cnotcount} \em{Variation in CNOT counts across different implementations with increasing input size.}}
	\end{figure}
\end{center}

We also present a bound on the CNOT counts of the considered implementations. In the proposed Toom-$2.5$ circuit shown in Fig.~\ref{fig:circuit}, CNOT gates are present in the Cuccaro adders and copy blocks. It can be seen from ~\citep{cuccaro2004new} that the number of CNOT gates in a $n$~bit adder can be bounded by $5n$. Proceeding similarly as the Toffoli count analysis, we get an exactly similar recurrence relation as presented in Gate Count Analysis in Section~\ref{sec:method}. Let $CC_n$ denote the number of CNOT gates in $T_{n,n}$. Also, let $Ac_n$ denote the number of CNOT for an in-place  $n$~bit adder. 
\begin{align}
 CC_n &= 16CC_{n/6} + 40Ac_{n/6} + 22Ac_{n/3} + 4Ac_{n/2} + 4Ac_{5n/6} \\
  &= 16^{log_6n}CC_1 + 40(Ac_{\frac{n}{6}} + 16Ac_{\frac{n}{36}}+\dots)  \nonumber \\ 
  &+ 22(Ac_{\frac{n}{3}} +  16Ac_{\frac{n}{18}}+\dots) +  4(Ac_{\frac{n}{2}} + 16Ac_{\frac{n}{12}}+\dots) \nonumber \\
   &+ 4(Ac_{\frac{5n}{6}} + 16Ac_{\frac{5n}{36}}+\dots) + COPY_{cnot}
\end{align}
where $COPY_{cnot}$ denotes the number of CNOTs used in the two copy blocks. However, the number of CNOT gates arising out of the Copy blocks are of the order $\mathcal{O}(n)$ and is dominated by the terms of order $n^{\log_{6}16}$. $CC_1 = 0$ because $1$-bit multiplier just consists of 1~Toffoli gate. 
\begin{align}
CC_n &\approx  2\big( 58n\big[\big( \frac{16}{6}\big)^{\log_6n}-1\big]\big)\\
&=116n( n^{\log_{6}(16/6)}-1)
\leq 116n^{\log_{6}16}
\end{align}
 By similar analysis,  the CNOT count of the Karatsuba multiplier can be bounded by $100n^{\log_{2}3}$. For the na\"ive method, controlled adders are considered as described in~\cite{parent2017improved}. Each such adder has $2n$~CNOTs and the multiplier uses~$n-1$ such adders. Thus, the total CNOT count is $2n^2-2n$. For the Constant multiplier in~\cite{pavlidis2012fast} $2n$~CNOT gates are employed. These observations are summarized in Fig.~\ref{fig:cnotcount}.

In~\cite{shende2008cnot}, it has been established that the $T$-gate is at least 6 times costlier compared to the CNOT gate, which emphasizes the importance of T-count and T-depth. However, with increasing circuit size, the cost of CNOT may take a dominant role if we follow the analysis in terms of upper/lower bounds~\cite{maslov2016optimal}. From that perspective, the study of overall cost is important. As we found for the case of multipliers, the CNOT count of Toom-$2.5$ Multiplier grows at slightly lower rate compared to that of the Karatsuba Multiplier with increasing input size. Considering the fact that Toffoli count of Toom-$2.5$ Multiplier already outperforms Karatsuba at large input sizes, the proposed design is clearly more efficient.

\section{\label{sec:conc}Conclusion}
\noindent Designing an efficient quantum circuit with low resource requirements and faster run time is an important challenge with significant repercussions across several domains, such as, scientific computing and security. In this work, we reported an efficient quantum circuit for integer multiplication based on Toom-Cook algorithm. We provide design results, and techniques for lowering the resource requirements. In terms of asymptotic complexity, the presented implementation outperforms the state-of-the-art results for multiple performance metrics.



\bibliography{apssamp}

\end{document}